\documentclass[conference]{IEEEtran}
\IEEEoverridecommandlockouts
\usepackage{booktabs}
\usepackage{cite}
\usepackage{comment}
\usepackage{amsmath,amssymb,amsfonts}
\usepackage{algorithmic}
\usepackage{graphicx}
\usepackage{textcomp}
\usepackage{xcolor}
\usepackage{hyperref}
\usepackage{multirow}
\usepackage{graphicx}
\usepackage{subfig}
\usepackage[linesnumbered,ruled,vlined]{algorithm2e}
\usepackage{tikz}
\def\BibTeX{{\rm B\kern-.05em{\sc i\kern-.025em b}\kern-.08em
    T\kern-.1667em\lower.7ex\hbox{E}\kern-.125emX}}
\usepackage{balance}

\hypersetup{
  colorlinks=true,
  linkcolor=blue,
  citecolor=blue,
  filecolor=magenta,
  urlcolor=blue
}

\def\BibTeX{{\rm B\kern-.05em{\sc i\kern-.025em b}\kern-.08em
    T\kern-.1667em\lower.7ex\hbox{E}\kern-.125emX}}
\begin{document}

\title{Performance Benchmarking of Machine Learning Models for Terahertz Metamaterial Absorber Prediction\\
}

% \and
% \and

\author{\IEEEauthorblockN{ Nafisa Anjum\IEEEauthorrefmark{1}, and Robiul Hasan \IEEEauthorrefmark{2}}
\IEEEauthorblockA{\IEEEauthorrefmark{1}\IEEEauthorrefmark{2}Dept. of Electrical \& Electronic Engineering (EEE), \\Rajshahi University of Engineering \& Technology, Bangladesh}
\IEEEauthorblockA{Emails: nafisaanjum9999@gmail.com, robiulhasan528527@gmail.com}
}

\maketitle

\begin{abstract}
This study presents a polarization-insensitive ultra-broadband terahertz metamaterial absorber based on vanadium dioxide (VO\textsubscript{2}) and evaluates machine learning methods for predicting its absorption performance. The structure consists of a VO\textsubscript{2} metasurface, a MF\textsubscript{2} dielectric spacer, and a gold ground plane. It achieves more than 90\% absorption between 5.72 and 11.11 THz, covering a bandwidth of 5.38 THz  with an average absorptance of 98.15\%. A dataset of 9,018 samples was generated from full-wave simulations by varying patch width, dielectric thickness, and frequency. Six regression models were trained: Linear Regression, Support Vector Regression, Decision Tree, Random Forest, XGBoost, and Bagging. Performance was measured using adjusted R², MAE, MSE, and RMSE. Ensemble models achieved the best results, with Bagging reaching an adjusted R² of 0.9985 and RMSE of 0.0146. The workflow offers a faster alternative to exhaustive simulations and can be applied to other metamaterial designs, enabling efficient evaluation and optimization.
\end{abstract}


\section{Introduction}
Terahertz (THz) metamaterial absorbers are artificial structures designed to achieve near-perfect absorption in the THz frequency range. They are ultra-thin, often sub-wavelength, and use resonant unit cells to trap and dissipate energy, enabling compact integration into devices beyond the limits of conventional absorbers.~\cite{chen2022terahertz}.

MMAs are used in THz technologies such as communication, imaging, sensing, and security due to their compact size and tunable response~\cite{liu2018ultra}. Since the first “perfect” absorber in 2008, many designs from single- to broadband have been developed~\cite{jain2024comparative}. Tunable MMAs using materials like VO\textsubscript{2} or graphene have gained attention for their ability to dynamically control absorption~\cite{chen2022terahertz, yang2022tunable}.

However, designing and optimizing MMAs remains a computationally intensive task. Traditional approaches rely on full-wave electromagnetic simulations, such as finite-element or time-domain solvers, which require significant resources for evaluating each candidate design. Exhaustive parameter sweeps and brute-force optimizations are especially time-consuming~\cite{jain2023machine, joy2025surrogate, soni2023machine}, which severely limits rapid exploration of high-performance absorber designs. For instance, optimizing a broadband or multi-resonant MMA could require simulating thousands of geometric variations, which becomes computationally infeasible~\cite{chamundeswari2024machine}. These limitations highlight the need for alternative techniques to accelerate the MMA design process. Machine learning (ML) has emerged as a promising solution to address these challenges in metamaterial research. By training ML models on datasets obtained from full-wave simulations, it is possible to rapidly predict the absorption response for new parameter sets, thereby reducing the reliance on exhaustive simulations~\cite{cerniauskas2024machine, joy2025surrogate}. ML models, especially deep learning techniques, are capable of learning complex nonlinear relationships and uncovering patterns that are often difficult to capture using traditional physics driven methods~\cite{qian2025guidance, soni2023machine, ding2023design}.

These ML-based surrogate models can be used for fast parametric sweeps, real-time tuning, and even inverse design of metamaterials, significantly improving the efficiency of the design workflow~\cite{qian2025guidance}. While several studies have explored ML for predicting absorber behavior and aiding in design, a comprehensive comparison of different ML algorithms for predicting MMA performance in the THz regime is still lacking~\cite{jain2024comparative, soni2023machine, gao2024deep}.

\begin{figure*}[htbp]
\centerline{\includegraphics[width=\linewidth]{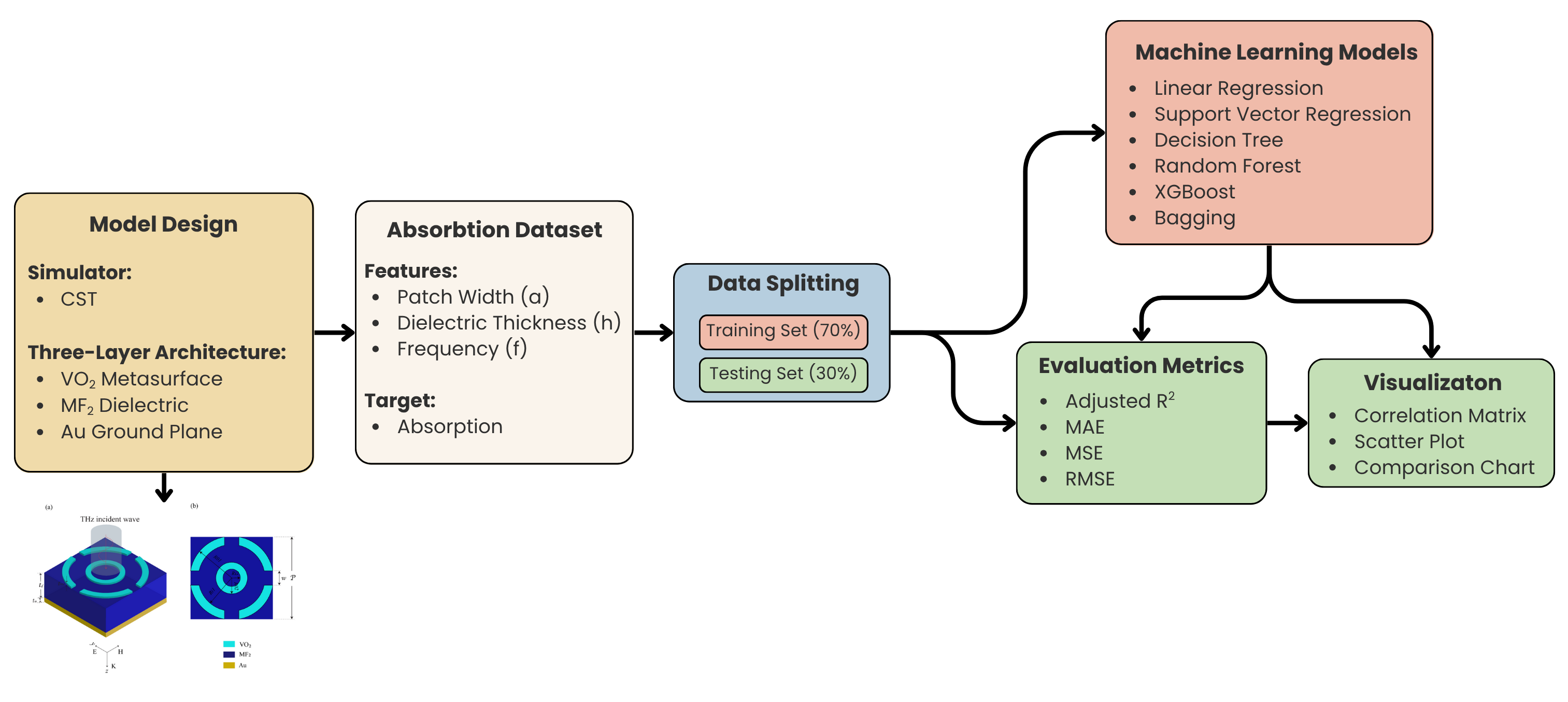}}
\caption{Flow chart illustrating the workflow of the proposed study.}
\label{fig:flowchart}
\end{figure*}

Motivated by these challenges, this work applies machine learning to predict the absorption behavior of a terahertz metamaterial absorber. A simulation-based dataset is generated from a CST model using ${\rm MF_2}$, ${\rm VO_2}$, and gold layers. The input features are geometric parameters $a$, $h$, and frequency $f$, and the target is absorptance. Six regression models are used: Linear Regression, SVR, Decision Tree, Random Forest, XGBoost, and Bagging. Model performance is evaluated using adjusted $R^2$, MAE, MSE, and RMSE. This study provides a side-by-side comparison of ML regressors for THz absorber modeling and offers practical guidance for data-driven metamaterial design.

\section{Literature Review}

\subsection{Metamaterial Absorbers in THz Regimes}

Metamaterials are engineered composite structures made up of subwavelength building blocks, commonly referred to as “meta-atoms,” which possess electromagnetic properties unattainable in natural materials.~\cite{qian2025guidance}. By adjusting the geometry and arrangement of these unit cells, metamaterials can be engineered to support effects such as negative refraction or tailored values of permittivity and permeability~\cite{qian2025guidance, jakvsic2006negative}. A key subclass, metamaterial absorbers (MMAs), uses resonant meta-atoms to absorb incident electromagnetic waves over specific frequency bands. Most MMAs follow a metal–dielectric–metal layered LC resonator configuration that traps incoming waves and dissipates energy as heat, resulting in near-perfect absorption at resonance~\cite{chen2022terahertz, landy2008perfect}.

The concept of a perfect absorber was first demonstrated in the microwave regime in 2008 and has since been extended to the THz domain~\cite{jain2024comparative}. In recent years, researchers have developed single-layer, multi-layer, and broadband MMAs, including tunable designs that incorporate materials such as semiconductors, graphene, or phase-change media to actively control absorption characteristics~\cite{shahsavaripour2025design}. While these advances highlight the potential of MMAs for THz wave manipulation, they also introduce a wide design space involving numerous geometric and material parameters, making manual optimization a difficult and time-consuming task~\cite{lee2024data}.

Conventional THz absorber design relies on iterative full-wave electromagnetic simulations, such as finite-difference time-domain (FDTD) or finite element method (FEM), to adjust resonator parameters and achieve the desired absorption characteristics~\cite{jain2024comparative}. While accurate, this approach is computationally demanding, as the simulation of a single 3D structure can require several hours. Evaluating thousands of design variations quickly becomes infeasible, especially when aiming to optimize broadband or multi-resonant absorbers. Additionally, traditional parametric sweeps often vary only a limited set of parameters, which can result in suboptimal performance by missing better design combinations. These challenges have motivated the search for faster and more scalable design methods, with growing interest in leveraging machine learning and artificial intelligence to accelerate optimization and reduce the reliance on exhaustive simulations~\cite{cerniauskas2024machine}.

\subsection{Machine Learning for Metamaterial Design and Analysis}

Over the past few years, machine learning has become an effective tool for accelerating the design of metamaterial absorbers and predicting their performance. For example, Patel et al. (2022) developed an ultra-broadband solar metasurface absorber and used a regression model to estimate its absorption spectrum from structural parameters~\cite{patel2022ultra}.

Jain et al. (2023) compared several regression algorithms for microwave-band MMAs, designing an X-/Ku-band absorber and optimizing it using decision trees, k-nearest neighbors, random forests, gradient boosting methods such as LightGBM and XGBoost, among others. They found that ensemble models, especially the Extra Trees regressor, produced the highest prediction accuracy, with adjusted $R^2$ values close to 0.99~\cite{jain2023machine}. Similar findings were reported by Prince et al. (2024), who evaluated multiple regressors including KNN, Decision Tree, Random Forest, Extra Trees, XGBoost, Bagging, and CatBoost on THz MMA data. They also identified Extra Trees as the top-performing model based on $R^2$, MSE, MAE, and RMSE metrics~\cite{jain2024comparative}.

Hou et al. (2020) introduced a deep learning approach for inverse design, using a neural network to generate absorber structures that met target specifications~\cite{hou2020customized}. More recent works show that ML is becoming integral to metamaterial research, expanding from early spectral response predictors to physics-informed models, generative frameworks, and adaptive optimizers~\cite{shahsavaripour2025design}.

Building on these developments, this study benchmarks a range of ML regressors to identify the most effective methods for predicting THz MMA absorption and provides practical insight for choosing models in future absorber design tasks.

\section{Methodology}
This section describes the procedure used to evaluate machine learning models for predicting the absorption response of a terahertz metamaterial absorber. An overview of the entire process, from absorber design to model evaluation, is shown in Fig.~\ref{fig:flowchart}. The study begins with the design and simulation of the absorber which provides absorption characteristics across controlled configurations. Next, a dataset is constructed by varying geometric parameters \(a\) and \(h\), and frequency \(f\), and by recording the corresponding absorption. The prepared dataset is then used to train and validate a set of supervised regression models. Finally, model performance is quantified using adjusted \(R^{2}\), MAE, MSE, and RMSE, which enables a consistent comparison across models.

\subsection{Metamaterial Absorber Design}

\begin{figure}[htbp]
\centerline{\includegraphics[width=\linewidth]{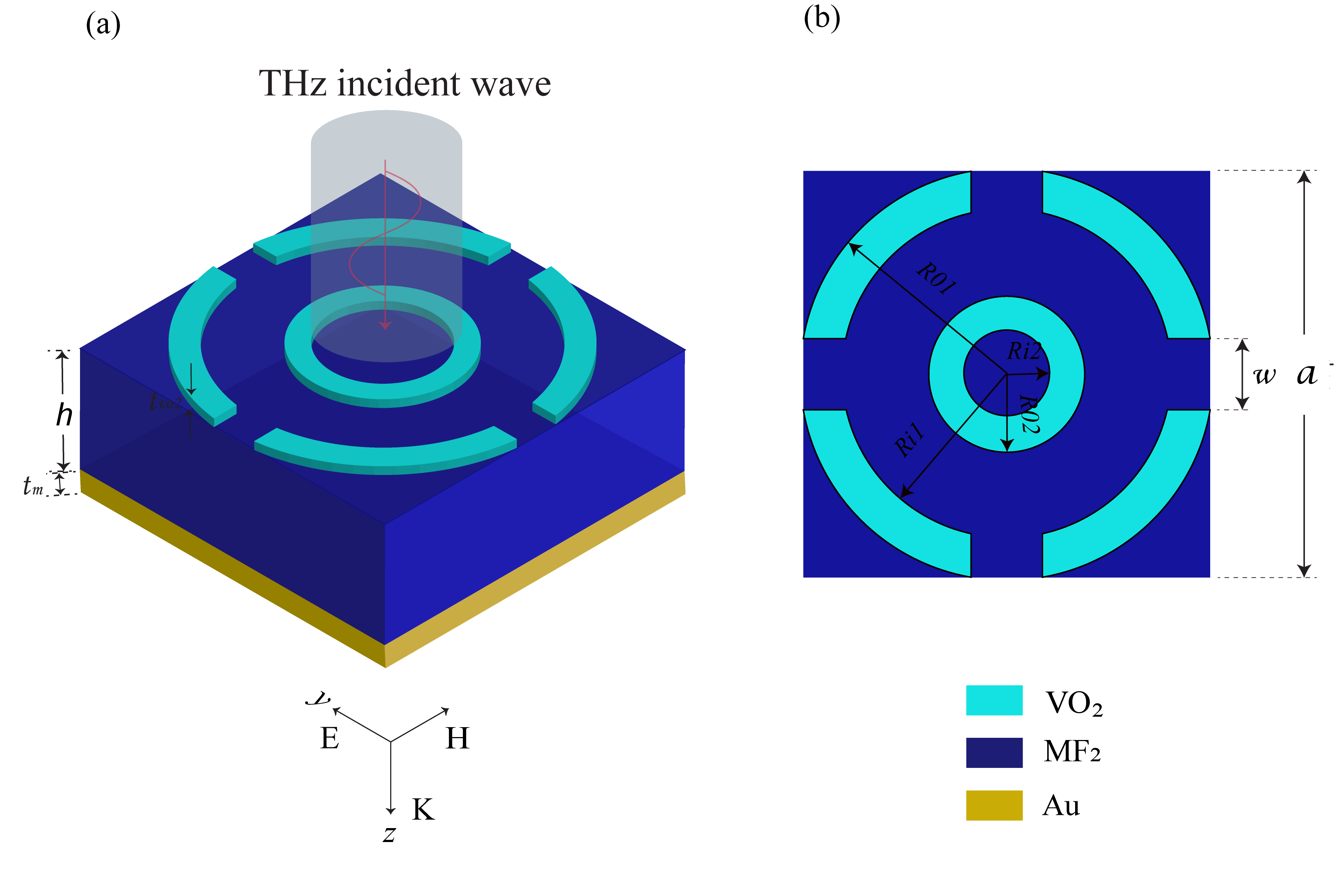}}
\caption{Geometry of the proposed metamaterial absorber: (a)  Three-dimensional schematic illustrating the layer arrangement and the direction of the incident THz wave; (b) Top-View Representation of Geometrical Parameters.}
\label{fig:model}
\end{figure}

\begin{figure*}[t]
\centering
% Row 1

\subfloat[Characteristics of Absorption and Reflection.]{\includegraphics[width=0.30\linewidth]{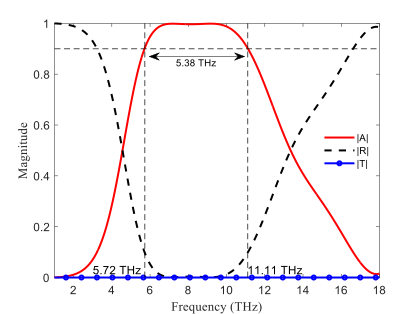}} \hfill
\subfloat[Absorption Spectra for Different Periodicities (a)]{\includegraphics[width=0.30\linewidth]{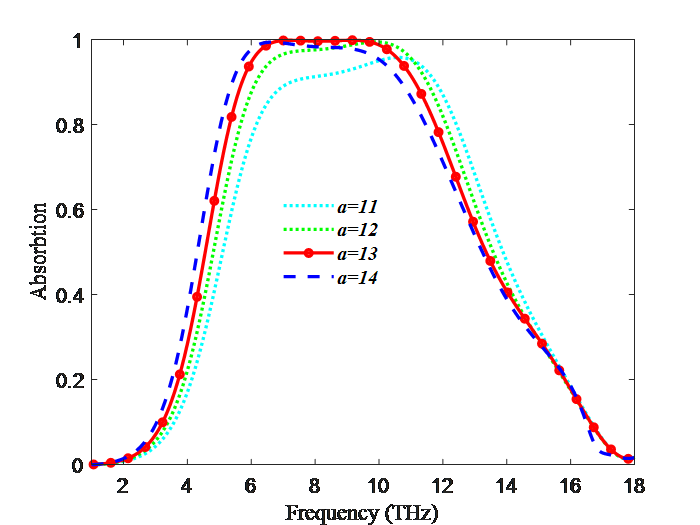}} \hfill
\subfloat[Absorption for different dielectric thickness(h)]{\includegraphics[width=0.30\linewidth]{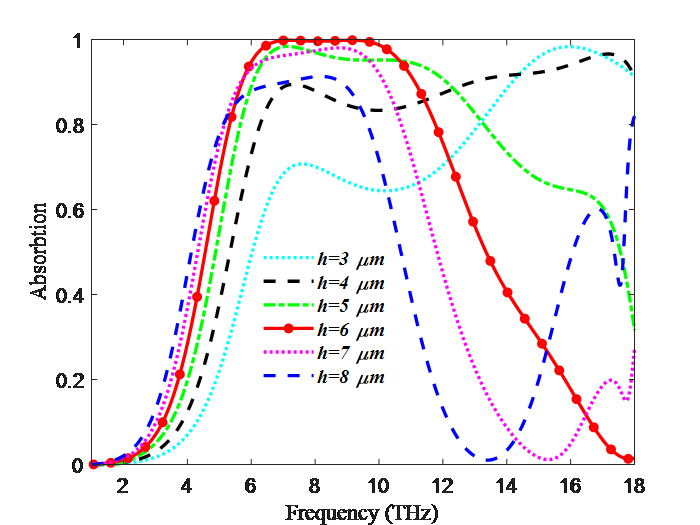}}

\caption{Simulated Absorption Response of the Proposed Metamaterial Absorber for Broadband Performance and Parameter Sensitivity Analysis}
\label{fig:simulation_results}
\end{figure*}

The absorber is engineered to achieve broadband terahertz absorption. As illustrated in Fig.~\ref{fig:model}, the uppermost layer consists of a patterned VO\textsubscript{2} metasurface that integrates a central closed ring along with two concentric split-ring resonators (SRRs), enabling the generation of multiple resonance modes. Beneath this, the middle layer is an MF\textsubscript{2} dielectric spacer with a relative permittivity of \(\varepsilon_r = 1.9\), which serves to match impedance. The bottom layer is a gold film, \(t_m = 0.02~\mu\text{m}\) thick, with an electrical conductivity of \(\sigma = 4.56 \times 10^7~\text{S/m}\), acting as a barrier to transmission while reflecting any residual waves. The VO$_2$ layer, with a thickness of $0.2~\mu\text{m}$, is described by the Drude model, characterized by a high-frequency dielectric constant $\varepsilon_{\infty} = 12$, a collision frequency of $\gamma = 5.75 \times 10^{13}~\text{rad/s}$, and a plasma frequency that varies according to its conductivity.

The unit cell has a periodicity of \(a = 13~\mu\text{m}\), a dielectric thickness \(h = 6~\mu\text{m}\), and the metal and VO\textsubscript{2} layers share the thicknesses specified above. The first ring has an outer radius of \(R_{o1} = 6.5~\mu\text{m}\) and an inner radius of \(R_{i1} = 5.5~\mu\text{m}\), with a strip width \(w = 1~\mu\text{m}\).The second ring features an outer radius of $R_{o2} = 2~\mu\text{m}$ and an inner radius of $R_{i2} = 1~\mu\text{m}$.

The unit cell is analyzed through finite element analysis (FEA), applying periodic boundary conditions along the $x$- and $y$-axes and a perfectly matched layer (PML) along the $z$-axis. Since the gold layer blocks transmission, the absorptance is determined by:
\begin{equation}
A(\omega) = 1 - \left|S_{11}(\omega)\right|^2
\end{equation}

\begin{table*}[ht]
\centering
\caption{Comparison of the Proposed Model with Recent Metamaterial Absorbers}
\resizebox{\textwidth}{!}{
\begin{tabular}{|p{1.3cm}|p{0.8cm}|c|c|p{6cm}|}
\hline
\textbf{Reference} & \textbf{Year} & \textbf{Bandwidth (THz)} & \textbf{Absorbance} & \textbf{Structural Design and Composition} \\
\hline
\multirow{2}{*}{\textbf{This Model}} & \multirow{2}{*}{} & \textbf{5.38} & \textbf{$>$90\%} & \multirow{2}{6cm}{\textbf{VO\textsubscript{2} based absorber that is polarization insensitive, ultra wideband, and high efficiency.}} \\[0.9em]
 &  & \textbf{3.35} & \textbf{$>$99\%} &  \\
\hline
 \cite{ge2025dual} & 2025 & 1.93 & $>90$\% & VO\textsubscript{2} microfluidic square resonator that works as both absorber and sensor, insensitive to polarization \\
\hline
\cite{ri2024tunable} & 2024 & 2.18 & $\sim$90\% & Graphene-integrated complementary split ring resonator with tunable properties.\\

\hline
 \cite{ryu2025theoretical} & 2025 & 4.70 & $>90$\% & VO\textsubscript{2} single resonator design that offers ultra wide bandwidth, polarization insensitivity, and high modulation depth. \\
 \hline
 \cite{zhang2024graphene} & 2024 & 2.5 & $\sim$92\% & Dual-Band Graphene Absorber with Independent Tuning.\\
\hline
\cite{xiong2025terahertz} & 2025 & 4.07 & $>95$\% & Hybrid VO\textsubscript{2} and graphene design with separate control of absorption and bandwidth. \\
\hline
\end{tabular}}
\label{table:comparison}
\end{table*}

\subsection{Dataset Collection}

The dataset was generated through parametric simulations using CST Microwave Studio by varying key design and operational parameters of the metamaterial absorber. Specifically, three variables were used as features: patch width (\(a\)) ranging from 11 to 14~\(\mu\)m, dielectric thickness (\(h\)) ranging from 3 to 8~\(\mu\)m, and frequency (\(f\)) ranging from 0 to 18~THz. 
% For each unique combination of these inputs, the corresponding absorptance value was computed using the relation \(A(\omega) = 1 - |S_{11}(\omega)|^2\), assuming zero transmittance due to the gold backing.
This process resulted in a total of 9,018 data points, each representing a unique configuration of the absorber. The resulting dataset was used as input for training and evaluating machine learning models aimed at predicting absorptance across the terahertz band.

\subsection{Machine Learning Models}

To predict the absorptance of the proposed metamaterial absorber from the input parameters \((a, h, f)\), six supervised regression models were implemented, representing a range of linear, nonlinear, and ensemble-based approaches. These include Linear Regression (LR) as a baseline linear method, Support Vector Regression (SVR) for capturing complex nonlinear relationships, Decision Tree Regressor (DT) as a non-parametric rule-based learner, Random Forest Regressor (RF) which averages multiple decision trees to reduce variance, Bagging Regressor as an ensemble approach using bootstrap sampling, and XGBoost Regressor, a high-performance gradient boosting method. Each model was trained and validated using the same dataset, with appropriate hyperparameter tuning applied where relevant. The objective was to evaluate and compare their predictive capabilities in terms of accuracy and generalization across the terahertz frequency range.

\subsection{Evaluation Metrics}

To evaluate the performance of the regression models, four commonly used statistical metrics were employed: adjusted \(R^2\), mean absolute error (MAE), mean squared error (MSE), and root mean squared error (RMSE). The adjusted \(R^2\) accounts for model complexity by penalizing the addition of irrelevant features. MAE measures the average magnitude of absolute errors, providing an intuitive sense of prediction accuracy. MSE emphasizes larger errors by squaring the deviations, while RMSE offers a scaled version of MSE with the same unit as the target variable. Together, these metrics provide a reliable basis for comparing the predictive performance and generalization ability of the models.

\section{Result and Discussion}

This section presents results from electromagnetic simulations of the proposed metamaterial absorber and from evaluating machine learning models for predicting its absorptance. Simulation results confirm broadband absorption and tunability, followed by a comparison of model performance using the metrics in Section~\ref{fig:flowchart}. The discussion covers model strengths, limitations, and implications for rapid absorber prediction.

\subsection{Metamaterial Absorber Simulation Results}

The proposed metamaterial absorber was designed and simulated to study its absorption behavior in the terahertz regime. The focus was on achieving a wide operating bandwidth, low reflection, and consistent performance under parameter variations.

Fig.~\ref{fig:simulation_results}(a) shows the simulated spectra for absorption, reflection, and transmission. The design maintains absorption above 90\% in the frequency range of 5.72~THz to 11.11~THz, giving a total bandwidth of 5.38~THz. Across this range, transmission is essentially zero due to the continuous metallic ground plane. Reflection remains minimal within the operational band and increases only outside the high-absorption region. The broadband performance is attributed to efficient impedance matching between the structure and free space, which reduces reflections and facilitates the absorption of incident energy.

The influence of the periodicity parameter $a$ on absorption is illustrated in Fig.~\ref{fig:simulation_results}(b). Variations in $a$ slightly shift the resonant frequency and bandwidth but preserve the broadband characteristic and high absorption level. This demonstrates that the absorber performance is robust against minor fabrication deviations in periodicity.

Fig.~\ref{fig:simulation_results}(c) illustrates the impact of modifying the dielectric layer thickness \(h\) on the absorption performance. While certain values of $h$ maintain the broadband high-absorption profile, others introduce secondary peaks or reduce absorption at higher frequencies. This behavior is due to changes in the effective impedance and resonance conditions within the structure. The selected $h$ value in the proposed design provides an optimal balance between strong broadband absorption and stable performance.

\subsection{Machine Learning Prediction Results}

\begin{table}[htbp]
\centering
\caption{Performance of Regression Models for Absorptance Prediction}
\label{tab:regression_results}
\begin{tabular}{lcccc}
\toprule
\textbf{Model} & \textbf{Adj. $R^2$} & \textbf{MAE} & \textbf{MSE} & \textbf{RMSE} \\
\midrule
Linear Regression & 0.1020 & 0.3191 & 0.1287 & 0.3587 \\
Support Vector Regression & 0.9828 & 0.0357 & 0.00247 & 0.0497 \\
Decision Tree Regressor & 0.9910 & 0.0229 & 0.00129 & 0.0359 \\
Random Forest Regressor & 0.9984 & 0.00834 & 0.000226 & 0.0150 \\
XGBoost Regressor & 0.9912 & 0.0221 & 0.00126 & 0.0355 \\
Bagging Regressor & \textbf{0.9985} & \textbf{0.00811} & \textbf{0.000212} & \textbf{0.0146} \\
\bottomrule
\end{tabular}
\end{table}

Table~\ref{tab:regression_results} presents the performance of six regression models for predicting absorptance from $(a, h, f)$. Five models achieved adjusted $R^2$ values above 0.98, confirming that the dataset and approach effectively capture the underlying patterns. Bagging Regressor outperformed all others, with an adjusted $R^2$ of 0.9985, MAE of 0.0081, MSE of 0.00021, and RMSE of 0.0146. Its ensemble strategy, averaging multiple base learners trained on bootstrap samples, reduces variance and improves prediction stability. Random Forest showed similar strengths, while Decision Tree performed reasonably well but was more sensitive to variance. XGBoost achieved good results but was slightly less accurate, likely because boosting is less effective on low-noise datasets. SVR captured nonlinear patterns but lagged behind the ensemble methods. Linear Regression was the weakest, with an adjusted $R^2$ of only 0.1020 and significantly higher errors, reflecting its inability to model the nonlinear input–output relationship.

\begin{figure}[htbp]
\centerline{\includegraphics[width=\linewidth]{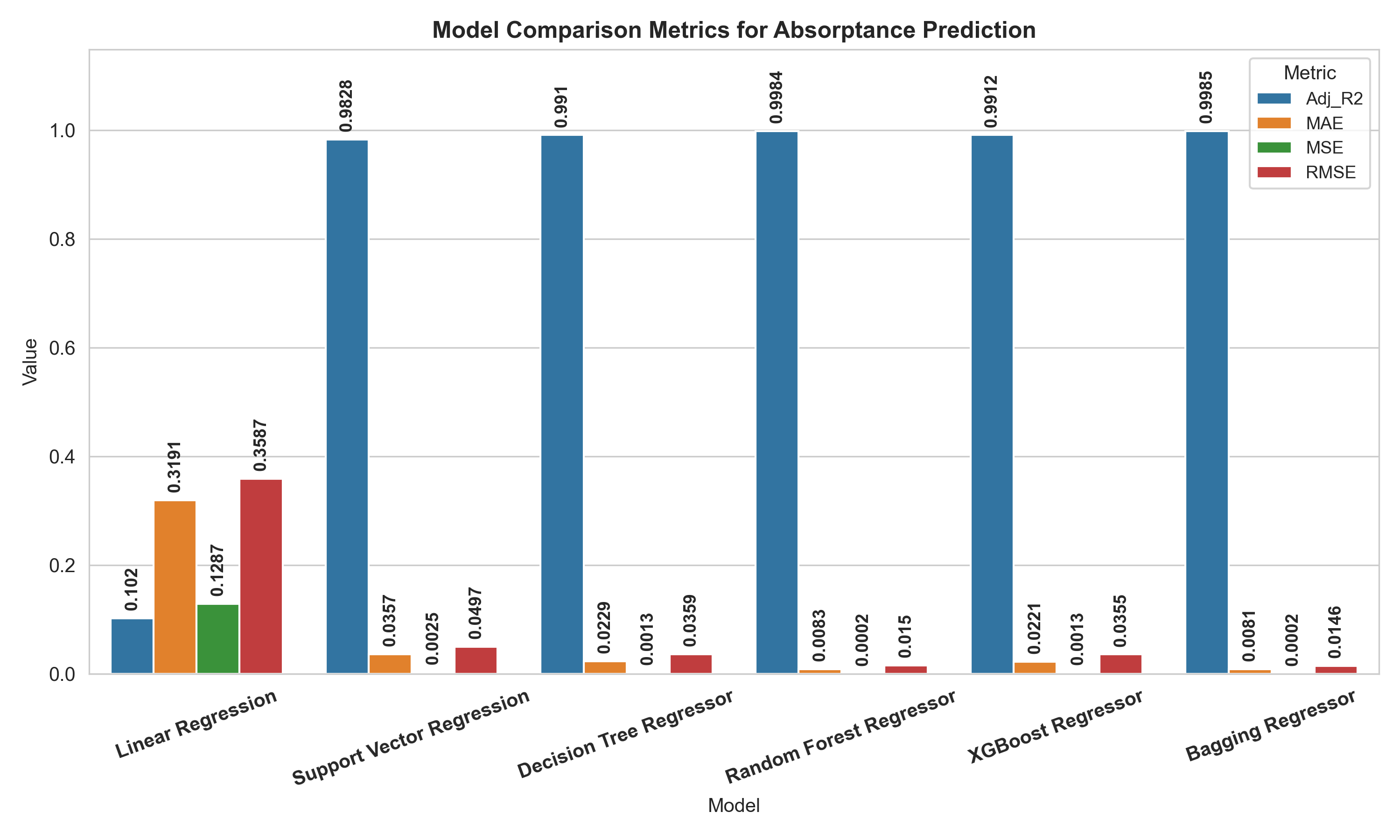}}
\caption{Performance comparison of six regression models for predicting absorptance using Adjusted R², MAE, MSE, and RMSE.}
\label{fig:model_comparison}
\end{figure}

Fig.~\ref{fig:model_comparison} visualizes the same trends. Bagging and Random Forest consistently yielded the highest $R^2$ and lowest error metrics. In contrast, Linear Regression stands out with the poorest alignment. Together, the table and figure highlight the robustness of ensemble-based models for this regression task.

\begin{figure}[htbp]
\centerline{\includegraphics[width=\linewidth]{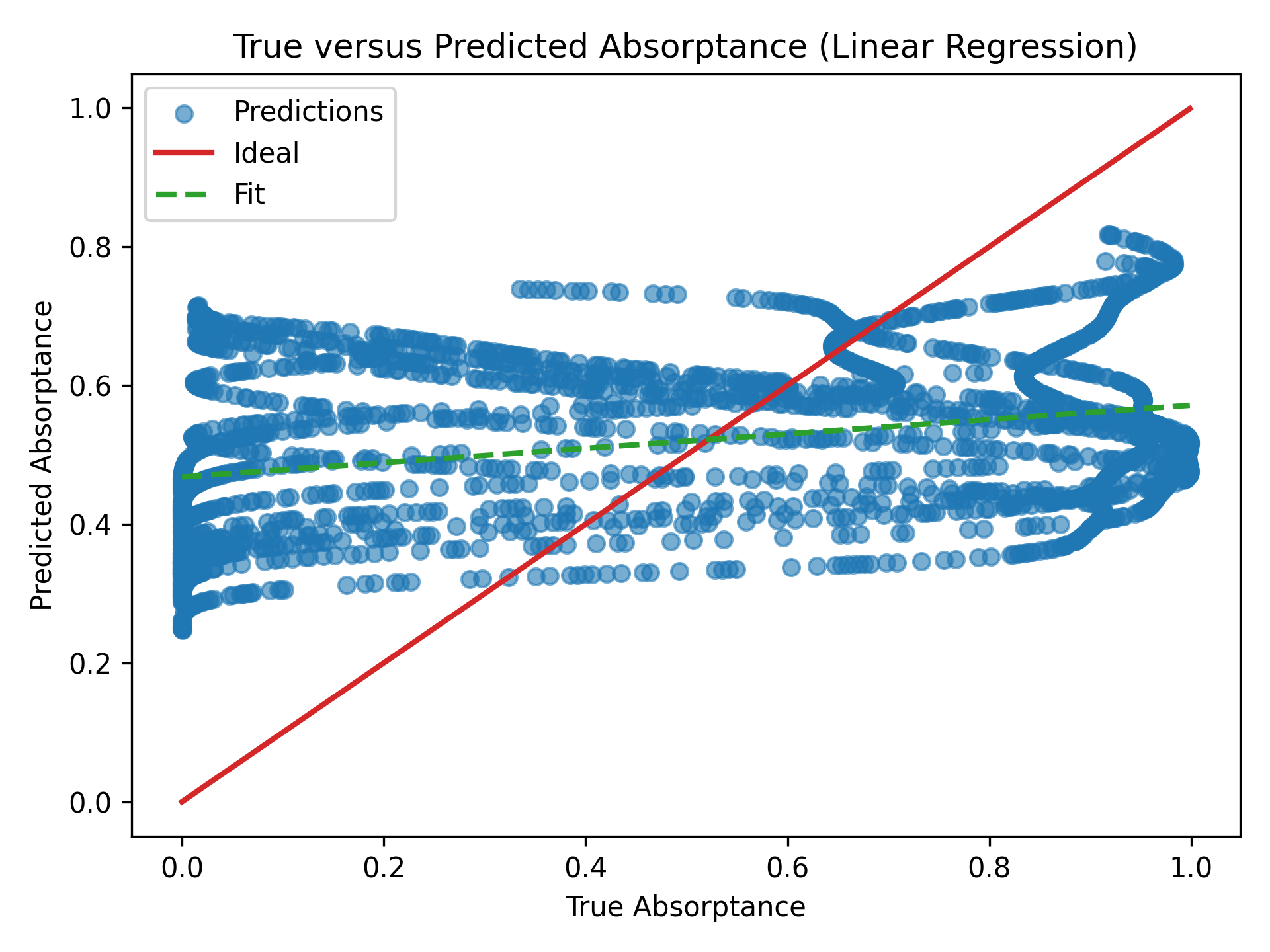}}
\caption{True versus predicted absorptance using the Linear Regression model}
\label{fig:linear}
\end{figure}

\begin{figure}[htbp]
\centerline{\includegraphics[width=\linewidth]{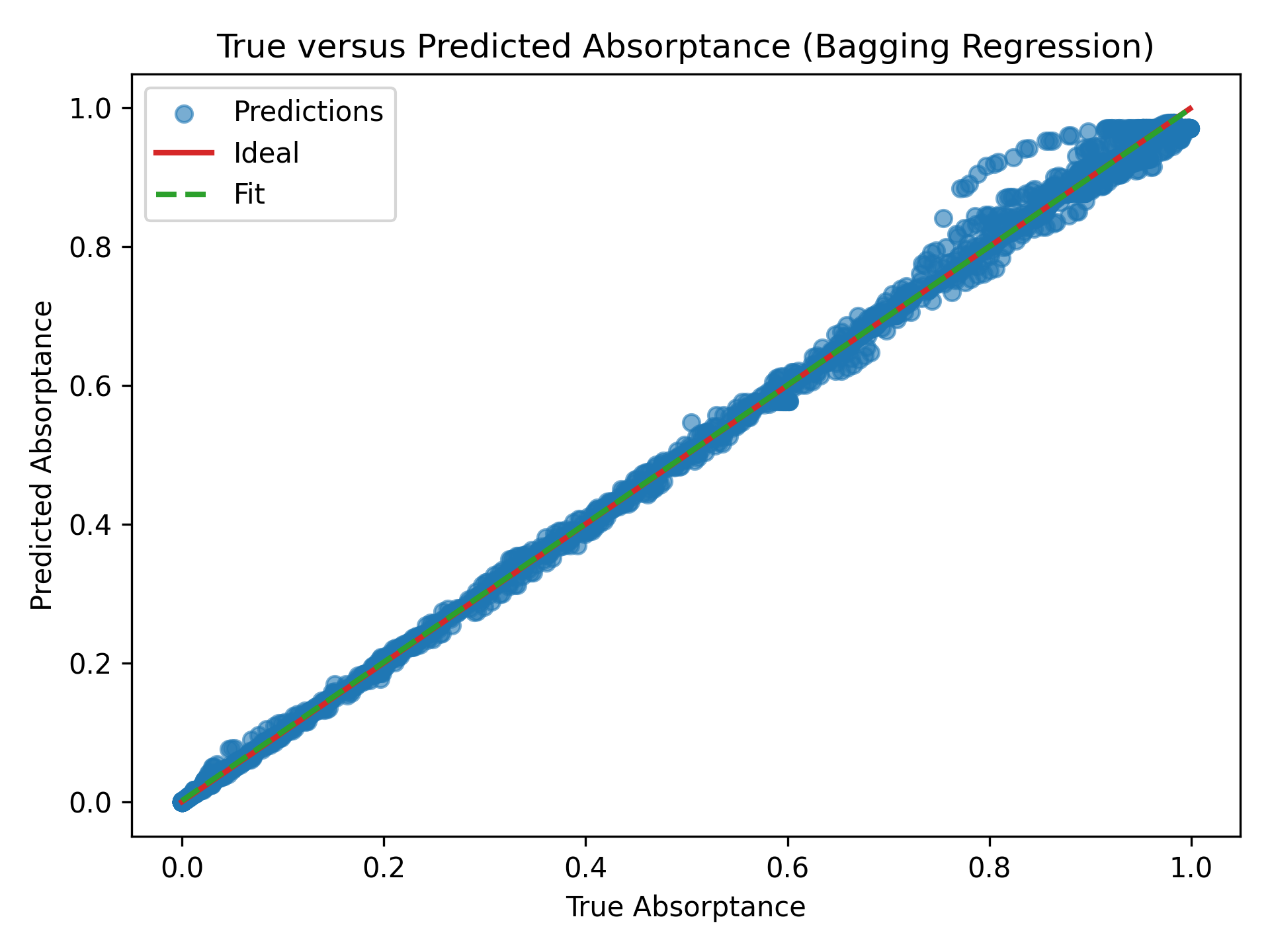}}
\caption{True versus predicted absorptance using the Linear Regression model}
\label{fig:bagging}
\end{figure}

Fig.~\ref{fig:linear} and Fig.~\ref{fig:bagging} compare the predicted versus true absorptance for the Linear Regression and Bagging Regression models. In Fig.~\ref{fig:linear}, the predictions from Linear Regression are widely scattered and poorly aligned with the red ideal line, which represents perfect agreement. The green trend line remains nearly flat, highlighting the model's inability to capture the nonlinear relationship between inputs and absorptance. This is expected, as Linear Regression assumes a strictly linear mapping, which does not suit the complex behavior of metamaterial absorbers.

By contrast, Fig.~\ref{fig:bagging} shows that the Bagging Regressor yields highly accurate predictions. The scatter points align closely with the diagonal, and the fitted line nearly overlaps the ideal line. Bagging’s ensemble of decision trees, each trained on different bootstrap samples, reduces variance and captures nonlinear patterns effectively. This makes it far more suitable for modeling the complex input–output relationships present in the absorber dataset, offering a clear advantage over simple regression methods.

\subsection{Summary of Findings}
In summary, the absorber maintains over 90\% absorptance across a 5.38~THz bandwidth due to effective impedance matching with low reflection and near zero transmission. The design is robust to changes in periodicity ($a$) and moderately sensitive to dielectric thickness ($h$). Machine learning models trained on this dataset showed high predictive accuracy. Bagging Regressor achieved the best performance, with an adjusted $R^2$ of 0.9985 and the lowest error across all metrics. Random Forest, XGBoost, and Decision Tree also performed well, while SVR captured nonlinear trends but lagged behind ensemble models. Linear Regression performed poorly, confirming the nonlinear nature of the mapping. The strong performance of the nonlinear models is due to the characteristics of both the absorber and the dataset. The simulated absorption profile is smooth and predictable, and the dataset is noise-free and continuous since it is generated from simulations. As a result, small changes in the features lead to consistent and learnable shifts in absorptance. 

\section{Data and Code Availability}
The dataset used in this study is publicly available at \href{https://www.kaggle.com/datasets/nanjum27/metamaterial-absorber-dataset}{Kaggle}. 
The implementation code for the machine learning models can be accessed at \href{https://github.com/Nafisa-21/Metamaterial-Machine-Learning-Implementation/tree/main}{GitHub}.

\section{Conclusion}

This study shows that machine learning is a reliable and efficient tool for predicting the absorptance of terahertz metamaterial absorbers. Using a simulation-driven dataset and six regression models, the research presents a practical workflow that reduces reliance on time-consuming full-wave simulations and accelerates design evaluation. The results highlight that ensemble models like Bagging and Random Forest achieved the highest accuracy, with Bagging reaching an adjusted $R^2$ of 0.9985. Most models performed well, confirming the dataset's suitability. Linear Regression underperformed, reflecting the nonlinear nature of the problem. Although the study focused on a single absorber configuration, the methodology can generalize to other metamaterial designs and related physical systems. These results show that ML-assisted modeling is both effective and scalable for similar engineering problems.
The main contribution of this work is a benchmarking framework that compares multiple ML regressors on a terahertz absorber dataset. It also introduces a repeatable design to prediction process that can support other simulation-intensive applications.
Future work can explore inverse design, include uncertainty modeling, and validate ML predictions against experimental data. These extensions will improve reliability and real-world applicability. Practitioners working on absorber design are encouraged to incorporate machine learning into their workflow. Doing so can reduce development time, support rapid evaluation, and open up wider design spaces across the terahertz regime.

\bibliography{references}

\end{document}